# The orbital and superhump periods of the SU UMa-type dwarf nova V1212Tauri

Jeremy Shears, Tom Krajci, Enrique de Miguel, Ian Miller, Etienne Morelle, George Roberts, Richard Sabo, William Stein and Raoul Behrend

**Abstract**

We report CCD photometry of the superoutburst of the dwarf nova V1212 Tau obtained during 2011 January and February. The outburst amplitude was at least 6 magnitudes and it lasted at least 12 days. Three distinct superhump regimes were observed. Initially low amplitude superhumps (0.03 to 0.05 magnitude peak-to-peak) with $P_{sh}$ = 0.0782(52) d were present. The superhumps reached a maximum amplitude of 0.31 magnitudes at the beginning of the plateau phase, with $P_{sh}$ = 0.07031(96) d. Subsequently the star began to fade slowly. During the first part of the decline, the period increased with $dP_{sh}/dt$ = +1.62(9) x $10^{-3}$ and the amplitude of the superhumps also declined. Mid way through the slow decline, the superhumps partially regrew and this point coincided with a change to a new superhump regime during which the period decreased with $dP_{sh}/dt$ = -1.50(39) x $10^{-3}$. We determined the orbital period as $P_{orb}$ = 0.06818(64) d and the superhump period excess as ε = 0.034(15)

**Introduction**

V1212 Tau was initially identified by Parsamian *et al.* (1) in search for flare stars in the vicinity of M45 using the 40 inch Schmidt telescope at the Byurakan Astrophysical Observatory in Armenia. Flares were observed in 1970, 1972 and 1977. A search of archival photographic plates by Parsamian *et al.* (2) between 1947 and 1987 revealed 12 more flares, with a recurrence time of about 330 days and an amplitude of at least 6 magnitudes. Their subsequent investigations showed that rather than being a flare star, V1212 Tau is a dwarf nova of the SU UMa family. They noted that the object is not present on Palomar Observatory prints to a limiting magnitude of 21.5 (blue photographic).

The Catalina Real-Time Transient Survey (3) (CRTS) detected 4 outbursts of V1212 Tau between 2004 Jan and 2011 Jan: in 2004 Nov (V=16.4), 2005 March (V=17.2), 2006 Sep (V=16.1) and 2007 Jan (V=16.8). In quiescence the object is below the CRTS detection limit with V > 20.5. The AAVSO International Database reveals two further outbursts in 2007 Nov (C=15.5; C= CCD with Clear filter) and 2009 Apr (V=16.2) in addition to the one reported in this paper. It is possible that other outbursts have been missed, not least because of the large seasonal gap that results from the object being located close to the ecliptic. Considering the interval between these outbursts, the shortest of which is about 4 months, it is possible that V1212 Tau outbursts more often than suggested by Parsamian et al. (2).

In this paper we present the results of a photometric campaign on the superoutburst of V1212 Tau observed during 2011 Jan and Feb. This intensive campaign was conducted by a worldwide team of astronomers using small telescopes (0.28 to 0.43 m) equipped with CCD cameras.

**Photometry and analysis**

Approximately 140 hours of unfiltered photometry was conducted during the outburst of V1212 Tau using the instrumentation shown in Table 1 and according to the observation log in Table 2. Images were dark-subtracted and flat-fielded prior to being measured using differential aperture photometry relative to either GSC 1804-0721 (V = 12.32), GSC 1804-0721(V = 11.7) or GSC 1804-0686, (V = 14.81). Given that the observers used different comparison stars and instrumentation, including CCD cameras with different spectral responses, small systematic differences are likely to exist between observers. However, given that the main aim of our research was to look for time dependent phenomena, we do not consider this to be a significant disadvantage. Nevertheless, where overlapping datasets were obtained during the outburst, we aligned measurements by different observers by experiment. Adjustments of up to 0.1 magnitudes were made. Heliocentric corrections were applied to all data.

**Detection and course of the 2011 Feb outburst**

The outburst was detected by JS (4) on 2011 Jan 27 at mag 15.6 during the course of a search for infrequently outbursting dwarf novae. The overall light curve of the outburst is shown in the top panel of Figure 1 and expanded views of the longer photometry runs in Figure 2, where each panel shows 2 days of data drawn to the same scale. Photometry conducted later on discovery night revealed the star fading at 0.4 mag/d (Figure 2 a). The following night, the mean brightness had stabilised at about magnitude 15.7, where it remained for about 3 days (JD 2455590 to 2455592), representing a short plateau phase. It then gradually declined at 0.12 mag/d for the next 8 days. Thus the outburst lasted at least 12 days, although the decline to quiescence was missed. Taking the quiescence magnitude as >21.5 means that the outburst amplitude was > 6 mag, as reported by Parsamian et al. (2).

**Measurement of the superhump period**

The photometry in Figure 2 clearly shows the presence of regular modulations which we interpret as superhumps, confirming this to be a superoutburst. To study the superhump behaviour, we first extracted the times of each sufficiently well-defined superhump maximum by using the Kwee and van Woerden method (5) in the *Minima v2.3* software (6). Times of 73 superhump maxima were found and are listed in Table 3. An unweighted analysis of the times of maximum during the plateau phase between JD 2455590 and 2455592 (the first superhump on JD 2455590 was designated cycle 0) allowed us to obtain the following linear superhump maximum ephemeris:

$$HJD_{max} = + 2455590.33383(89) + 0.07031(96) \times E \qquad \text{Equation 1}$$

This gives the mean superhump period for this stage of the superoutburst $P_{sh}$ = 0.07031(96) d. The O–C residuals for the superhump maxima for the complete outburst relative to the ephemeris are shown in the middle panel in Figure 1.

**Superhump evolution**

The O-C diagram in Figure 1 (middle panel) shows that the superhump period changed significantly during the outburst. Kato *et al.* (7) studied the superhump period evolution in a large number of SU UMa systems and found that many outbursts appeared to show three distinct stages: an early evolutionary stage (A) with longer superhump period, a middle stage (B) during which systems with $P_{orb}$ < 0.08 d have a positive period derivative, and a final stage (C) with a shorter $P_{sh}$. Our O-C diagram for V1212 Tau appears to be a classical stage A-B-C transition and is very similar to O-C diagrams of SW UMa and UV Per shown in Kato et al. (7), amongst many others which exhibit this evolutionary trend.

Stage A of the superoutburst, shown by the blue circles in Figure 1, was brief and constrained to the first 2 nights of the outburst (JD 2455589 and 2455590). The mean superhump period in the interval was $P_{sh}$ = 0.0731(12) d, or about 4% longer than that given in Equation 1. This is typical for SU UMa systems where $P_{sh}$ in the early stages of an outburst tends to be ~1 to 4% longer than the mean $P_{sh}$ later in the outburst (7).

The interval between JD 2455591 and 2455595 (red data points in Figure 1) corresponds to Stage B, during which we found an increasing superhump period with $dP_{sh}/dt$ = +1.62(9) x $10^{-3}$ by fitting a quadratic function to the data (red dotted line in Figure 1). The transition from stage A to B appears to coincide with the end of the brief plateau phase and the beginning of the slow decline.

Mid way through the slow decline, at around JD 2455595, there was a transition to stage C superhumps which persisted for the rest of the observed outburst. During this stage, the superhump period decreased with $dP_{sh}/dt$ = -1.50(39) x $10^{-3}$ as shown by the quadratic fit to the data between JD 2455595 and JD 2455601 (green trend line in Figure 1).

We also found that the superhump peak-to-peak amplitude changed significantly during the outburst (Figure 1, bottom panel) and appeared to correlate with the three stages of the outburst. During stage A, the amplitude gradually increased from 0.03 to 0.17 magnitudes. The transition to stage B coincided with a maximum superhump amplitude of 0.31 magnitudes, followed by a decline to about 0.16 magnitudes. At the start of stage C, the superhumps regrew briefly to 0.19 magnitudes and subsequently declined during the rest of the outburst, with the last observed superhump being 0.07 magnitudes. A very similar evolutionary trend in superhump amplitude was observed in SW UMa (7) (8) where the maximum amplitude coincided with the transition from stage A to B and a secondary maximum occurred at the transition from stage B to C.

**Determining the orbital period and estimation of the binary mass ratio**

Close inspection of the light curves revealed that the superhump profiles were not always smooth. In some SU UMa systems irregularities, or humps superimposed on the underlying superhump profile, have been observed and shown to be orbital modulations from which the orbital period can be determined (9). We performed a Lomb-Scargle analysis on the data from JD 2455590 and 2455591, having subtracted the mean magnitude, using the *Peranso* software version 2.50 (10). We chose not to include the photometry from the other nights since, as the value of $P_{sh}$ was evolving, the resulting superhump signal in the power spectrum broadened and thus it was not possible to remove the signal completely by pre-whitening. The resulting power spectrum is shown in Figure 3a which has its highest peak at 14.182(69) cycles/d, and its 1 cycle/d aliases, which we interpret as the superhump signal with $P_{sh}$ = 0.07051(34) d. The error estimates are derived using the Schwarzenberg-Czerny method (11). Pre-whitening the power spectrum with the superhump signal gave the Lomb-Scargle spectrum in Figure 3b. Here the most powerful signal was at 14.667(139) cycles/d (and its 1 cycle/d aliases), which we interpret as the orbital signal, corresponding to $P_{orb}$ = 0.06818(64) d. To our knowledge, this is the first direct measurement of $P_{orb}$ in V1212 Tau (an estimate of $P_{orb}$ = 0.073 d is given in Ritter & Kolb Catalogue of Cataclysmic Variables, edition 7.15 (12), but this was estimated from an earlier measurement of $P_{sh}$ using an empirical relation between $P_{sh}$ and $P_{orb}$).

Using our values of $P_{sh}$ = 0.07051(34) d and $P_{orb}$ = 0.06818(64) d allows the fractional superhump period excess, $\varepsilon = (P_{sh} - P_{orb})/ P_{orb}$, to be calculated as $\varepsilon$ = 0.034(15). This value is consistent with the range of $\varepsilon$ observed in other SU UMa dwarf novae having similar $P_{orb}$ (13). Measuring $\varepsilon$ provides a way to estimate the mass ratio, q= $M_{sec}/M_{wd}$, of a CV and following Patterson *et al.* (13) we find q $\approx$ 0.15 for V1212 Tau.

**Discussion**

Although our observations are consistent with a classical stage A-B-C superoutburst transition as observed in many SU UMa systems (7), one aspect was unusual: the fading trend during stage A. Such a trend is reminiscent of the normal outbursts that have been reported in several SU UMa systems prior to the start of the superoutburst itself. These precursor outbursts, which are often modulated with the orbital period, have been proposed as triggers for the superoutburst (15). To investigate the 0.03 to 0.05 magnitude modulations observed on the first night (JD2455589) further, and to test our interpretation that they were stage A superhumps rather than orbital modulations, we performed a Lomb-Scargle analysis on the photometry, having first removed the linear fading trend. The resulting power spectrum is shown in Figure 4 which has its highest peak at 12.780(849) cycles/d, or P = 0.0782(52) d. Such a period is 3.1% longer than the $P_{orb}$. Thus we confirm our interpretation that these modulations are superhumps, with $P_{sh}$ = 0.0782(52) d.

The presence of superhumps during this period in V1212 Tau means that the superoutburst had only just begun and the accretion disc had already extended to the 3:1 resonance radius. A similar situation was observed during the 1993 outburst of T Leo (= QZ Vir) where there was a precursor before the main superoutburst when the star declined at 0.84 mag/d and during which superhumps were present, the period of which was 2.4% longer than $P_{orb}$ (16). By contrast, observations of V342 Cam showed that orbital humps were present during the fade from a precursor outburst. Superhumps with an unusually long period developed subsequently during the rise to the superoutburst maximum (9).

We suggest that a detailed analysis of precursor outbursts and the early stages of superoutbursts in other SU UMa systems would be useful, specifically regarding the presence of orbital humps and superhumps in these stages and whether the system was brightening or fading. No global study has yet been carried out on the early stages of superoutbursts, partly because this stage is short-lived and often lacks observational coverage, and it might throw light on the cause of superoutbursts.

**Conclusions**

An intensive CCD photometry campaign was conducted during the superoutburst of V1212 Tau in 2011 January and February. This showed that the outburst amplitude was at least 6 magnitudes and that it lasted at least 12 days. Three distinct superhump regimes were observed. In stage A, at the beginning of the outburst, low amplitude superhumps were present (0.03 to 0.05 magnitude peak-to-peak) with $P_{sh}$ = 0.0782(52) d. The superhumps reached a maximum amplitude of 0.31 magnitudes at the beginning of the plateau phase, with $P_{sh}$ = 0.07031(96) d. This coincided with the start of stage B during which the star began to fade slowly and the superhump period increased with $dP_{sh}/dt$ = +1.62(9) x $10^{-3}$. The amplitude of the superhumps also declined during this stage. Mid way through the slow decline, the superhumps partially regrew and this point coincided with a change to a new superhump regime, stage C, during which the period decreased with $dP_{sh}/dt$ = -1.50(39) x $10^{-3}$. We determined the orbital period as $P_{orb}$ = 0.06818(64) d and the superhump period excess was ε = 0.034(15), which is consistent with other SU UMa systems of similar $P_{orb}$.

**Acknowledgements**

The authors gratefully acknowledge the use of data from the Catalina Real-Time Transient Survey, kindly provided to us by Dr. Andrew Drake, and data from the AAVSO International Database contributed by observers worldwide. This research made use of SIMBAD, operated through the Centre de Données Astronomiques (Strasbourg, France) and the NASA/Smithsonian Astrophysics Data System. We thank our referees, Dr. Coel Hellier and Dr. Tom Lloyd Evans, for their helpful comments.

**Addresses**

JS: "Pemberton", School Lane, Bunbury, Tarporley, Cheshire, CW6 9NR, UK [bunburyobservatory@hotmail.com]

TK: CBA New Mexico, PO Box 1351 Cloudcroft, New Mexico 88317, USA [tom_krajci@tularosa.net]

EdM: Departamento de Fisica Aplicada, Facultad de Ciencias Experimentales, Universidad de Huelva, 21071 Huelva, Spain; Center for Backyard Astrophysics, Observatorio del CIECEM, Parque Dunar, Matalascañas, 21760 Almonte, Huelva, Spain [demiguel@uhu.es]

IM: Furzehill House, Ilston, Swansea, SA2 7LE, UK [furzehillobservatory@hotmail.com]

EM: Lauwin-Planque Observatory, F-59553 Lauwin-Planque, France [etmor@free.fr]

GR: 2007 Cedarmont Dr., Franklin, TN 37067, USA, [georgeroberts0804@att.net]

RS: 2336 Trailcrest Dr., Bozeman, MT 59718, USA [richard@theglobal.net]

WS: 6025 Calle Paraiso, Las Cruces, NM 88012, USA [starman@tbelc.org]

RB: Geneva Observatory, CH-1290 Sauverny, Switzerland [raoul.behrend@unige.ch]

**References**

1. *Parsamian E.A. et al., Astronomicheskii Tsirkulyar, (1983).*

2. *Parsamian E.A., Gonzales G. and Ohanian G.B., Astrophysics, 42, 33 (1999).*

3. *Drake A.J. et al., ApJ, 696, 870 (2009).*

4. *Shears J., http://tech.groups.yahoo.com/group/baavss-alert/message/2473.*

5. *Kwee K.K. and van Woerden H., Bull. Astron. Inst. Netherlands, 12, 327-330 (1956).*

6. *Nelson R. (2007) www.members.shaw.ca/bob.nelson/software1.htm.*

7. *Kato T. et al., PASJ,61, S395-S616 (2009).*

8. *Soejima Y. et al., PASJ, 61, 659-674 (2009) .*

9. *Shears J.H., New Astronomy, 16, 311-316 (2011).*

10. *Vanmunster T., Peranso, http://www.peranso.com.*

11. *Schwarzenberg-Czerny A., MNRAS., 253, 198 (1991).*

12. *Ritter H. and Kolb U., A&A, 404, 301 (2003).*

13. *Patterson J. et al., PASP, 117, 1204-1222 (2005).*

14. *Marino B.F. and Walker W.S.G. in Changing Trends in Variable Star Research, IAU Colloquium 46, F.M. Bateson, J. Smak and J.H. Urch eds., University of Waikato, Hamilton, NZ, pg 29 (1979).*

15. *Kato T., PASJ, 49, 583-587 (1997).*

| Observer | Telescope | CCD |
|---|---|---|
| Krajci | 0.3 m SCT | SBIG ST9-XME |
| de Miguel | 0.28 m SCT | QSI-516ws |
| Miller | 0.35 m SCT | Starlight Xpress SXVF-H16 |
| Morelle [a] | 0.4 m SCT | SBIG ST-9 |
| Roberts | 0.4 m SCT | SBIG ST-8 |
| Sabo | 0.43 m reflector | SBIG STL-1001 |
| Shears | 0.28 m SCT | Starlight Xpress SXVF-H9 |
| Stein | 0.35 m SCT | SBIG ST-10XME |

[a] photometric reduction by Raoul Behrend

**Table 1: Instrumentation**

| Start date in 2011 (UT) | Start time (JD) | End time (JD) | Duration (h) | Observer |
|---|---|---|---|---|
| Jan 28 | 2455589.574 | 2455589.793 | 5.3 | Krajci |
| Jan 28 | 2455590.267 | 2455590.485 | 5.2 | Shears |
| Jan 28 | 2455590.401 | 2455590.512 | 2.7 | de Miguel |
| Jan 28 | 2455590.439 | 2455590.554 | 2.8 | Miller |
| Jan 29 | 2455590.572 | 2455590.729 | 3.8 | Roberts |
| Jan 29 | 2455590.556 | 2455590.792 | 5.7 | Krajci |
| Jan 29 | 2455591.348 | 2455591.520 | 4.1 | de Miguel |
| Jan 29 | 2455591.396 | 2455591.543 | 3.5 | Miller |
| Jan 30 | 2455591.525 | 2455591.693 | 4.0 | Roberts |
| Jan 30 | 2455591.555 | 2455591.788 | 6.0 | Krajci |
| Jan 30 | 2455592.279 | 2455592.507 | 5.5 | de Miguel |
| Jan 30 | 2455592.392 | 2455592.507 | 2.8 | Miller |
| Jan 31 | 2455592.590 | 2455592.731 | 3.4 | Stein |
| Jan 31 | 2455593.299 | 2455593.481 | 4.4 | Morelle |
| Jan 31 | 2455593.325 | 2455593.497 | 4.1 | de Miguel |
| Feb 1 | 2455593.564 | 2455593.763 | 4.8 | Sabo |
| Feb 1 | 2455594.261 | 2455594.479 | 5.2 | Morelle |
| Feb 1 | 2455594.280 | 2455594.509 | 5.5 | de Miguel |
| Feb 2 | 2455594.608 | 2455594.811 | 4.8 | Sabo |
| Feb 2 | 2455595.281 | 2455595.500 | 5.3 | de Miguel |
| Feb 2 | 2455595.326 | 2455595.518 | 4.6 | Miller |
| Feb 2 | 2455595.337 | 2455595.475 | 3.3 | Morelle |
| Feb 3 | 2455595.636 | 2455595.731 | 2.3 | Sabo |
| Feb 3 | 2455596.303 | 2455596.499 | 4.7 | de Miguel |
| Feb 4 | 2455597.297 | 2455597.499 | 4.8 | de Miguel |
| Feb 4 | 2455597.321 | 2455597.482 | 3.9 | Morelle |
| Feb 5 | 2455598.273 | 2455598.479 | 4.9 | Morelle |
| Feb 5 | 2455598.323 | 2455598.497 | 4.2 | de Miguel |
| Feb 6 | 2455599.252 | 2455599.477 | 5.4 | Morelle |
| Feb 6 | 2455599.317 | 2455599.493 | 4.2 | de Miguel |
| Feb 7 | 2455600.274 | 2455600.474 | 4.8 | Morelle |
| Feb 8 | 2455601.274 | 2455601.402 | 3.1 | Morelle |
| Feb 9 | 2455602.300 | 2455602.359 | 1.4 | Morelle |

**Table 2: Time series observation log**

| Superhump cycle number | Superhump maximum (HJD) | Measurement error (d) | O-C (d) | Superhump amplitude (mag) |
|---|---|---|---|---|
| -10 | 2455589.5895 | 0.0057 | -0.0413 | 0.03 |
| -9 | 2455589.6634 | 0.0045 | -0.0378 | 0.04 |
| -8 | 2455589.7427 | 0.0033 | -0.0288 | 0.04 |
| 0 | 2455590.3228 | 0.0030 | -0.0112 | 0.12 |
| 1 | 2455590.4071 | 0.0024 | 0.0027 | 0.12 |
| 2 | 2455590.4726 | 0.0015 | -0.0021 | 0.11 |
| 2 | 2455590.4715 | 0.0009 | -0.0032 | 0.13 |
| 2 | 2455590.4704 | 0.0015 | -0.0043 | 0.12 |
| 3 | 2455590.5467 | 0.0012 | 0.0017 | 0.13 |
| 4 | 2455590.6176 | 0.0006 | 0.0023 | 0.17 |
| 4 | 2455590.6154 | 0.0015 | 0.0001 | 0.16 |
| 5 | 2455590.6891 | 0.0030 | 0.0035 | 0.18 |
| 5 | 2455590.6868 | 0.0015 | 0.0012 | 0.16 |
| 6 | 2455590.7538 | 0.0024 | -0.0022 | 0.17 |
| 15 | 2455591.3889 | 0.0006 | 0.0001 | 0.27 |
| 16 | 2455591.4628 | 0.0003 | 0.0036 | 0.26 |
| 16 | 2455591.4627 | 0.0006 | 0.0035 | 0.29 |
| 17 | 2455591.5335 | 0.0009 | 0.0040 | 0.30 |
| 17 | 2455591.5310 | 0.0006 | 0.0015 | 0.29 |
| 18 | 2455591.6010 | 0.0006 | 0.0012 | 0.31 |
| 18 | 2455591.6001 | 0.0009 | 0.0003 | 0.29 |
| 19 | 2455591.6691 | 0.0006 | -0.0010 | 0.29 |
| 19 | 2455591.6713 | 0.0006 | 0.0012 | 0.28 |
| 20 | 2455591.7419 | 0.0003 | 0.0015 | 0.28 |
| 28 | 2455592.3029 | 0.0009 | -0.0001 | 0.25 |
| 29 | 2455592.3694 | 0.0012 | -0.0039 | 0.25 |
| 30 | 2455592.4417 | 0.0009 | -0.0019 | 0.26 |
| 30 | 2455592.4418 | 0.0012 | -0.0018 | 0.24 |
| 33 | 2455592.6515 | 0.0006 | -0.0031 | 0.20 |
| 34 | 2455592.7208 | 0.0012 | -0.0041 | 0.21 |
| 43 | 2455593.3489 | 0.0012 | -0.0089 | 0.22 |
| 43 | 2455593.3499 | 0.0012 | -0.0079 | 0.18 |
| 44 | 2455593.4192 | 0.0009 | -0.0089 | 0.22 |
| 44 | 2455593.4201 | 0.0015 | -0.0080 | 0.21 |
| 45 | 2455593.4903 | 0.0015 | -0.0081 | 0.22 |
| 47 | 2455593.6297 | 0.0006 | -0.0094 | 0.19 |
| 48 | 2455593.6994 | 0.0015 | -0.0100 | 0.19 |
| 57 | 2455594.3297 | 0.0015 | -0.0126 | 0.19 |
| 57 | 2455594.3306 | 0.0009 | -0.0117 | 0.17 |
| 58 | 2455594.4012 | 0.0012 | -0.0114 | 0.20 |
| 58 | 2455594.4020 | 0.0012 | -0.0106 | 0.18 |
| 59 | 2455594.4735 | 0.0012 | -0.0094 | 0.19 |
| 59 | 2455594.4719 | 0.0018 | -0.0110 | 0.18 |
| 62 | 2455594.6827 | 0.0018 | -0.0112 | 0.16 |
| 63 | 2455594.7510 | 0.0015 | -0.0132 | 0.17 |
| 71 | 2455595.3091 | 0.0021 | -0.0177 | 0.16 |
| 72 | 2455595.3859 | 0.0018 | -0.0112 | 0.17 |
| 72 | 2455595.3840 | 0.0015 | -0.0131 | 0.15 |
| 72 | 2455595.3828 | 0.0018 | -0.0143 | 0.17 |
| 73 | 2455595.4564 | 0.0015 | -0.0110 | 0.17 |
| 73 | 2455595.4552 | 0.0024 | -0.0122 | 0.14 |
| 73 | 2455595.4557 | 0.0012 | -0.0117 | 0.16 |
| 75 | 2455595.5960 | 0.0024 | -0.0120 | 0.16 |
| 76 | 2455595.6680 | 0.0027 | -0.0104 | 0.17 |
| 86 | 2455596.3687 | 0.0036 | -0.0129 | 0.17 |
| 87 | 2455596.4355 | 0.0030 | -0.0164 | 0.19 |
| 100 | 2455597.3486 | 0.0063 | -0.0174 | 0.17 |
| 100 | 2455597.3482 | 0.0018 | -0.0178 | 0.18 |
| 101 | 2455597.4151 | 0.0069 | -0.0213 | 0.17 |
| 101 | 2455597.4184 | 0.0051 | -0.0180 | 0.19 |
| 114 | 2455598.3249 | 0.0021 | -0.0256 | 0.14 |
| 115 | 2455598.3966 | 0.0024 | -0.0242 | 0.14 |
| 115 | 2455598.3908 | 0.0039 | -0.0300 | 0.14 |
| 116 | 2455598.4653 | 0.0045 | -0.0259 | 0.14 |
| 116 | 2455598.4640 | 0.0036 | -0.0272 | 0.13 |
| 128 | 2455599.2979 | 0.0054 | -0.0371 | 0.16 |
| 129 | 2455599.3701 | 0.0060 | -0.0352 | 0.14 |
| 129 | 2455599.3659 | 0.0084 | -0.0394 | 0.14 |
| 130 | 2455599.4402 | 0.0057 | -0.0354 | 0.14 |
| 130 | 2455599.4425 | 0.0072 | -0.0331 | 0.13 |
| 143 | 2455600.3443 | 0.0063 | -0.0455 | 0.09 |
| 144 | 2455600.4159 | 0.0057 | -0.0442 | 0.09 |
| 157 | 2455601.3190 | 0.0093 | -0.0553 | 0.07 |

**Table 3: Times and amplitudes of superhumps**

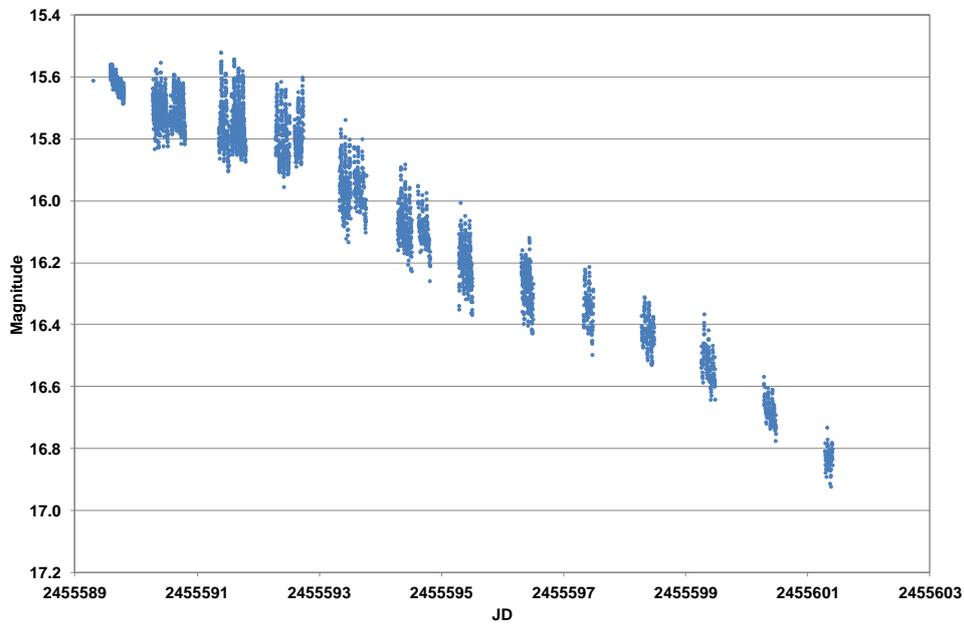
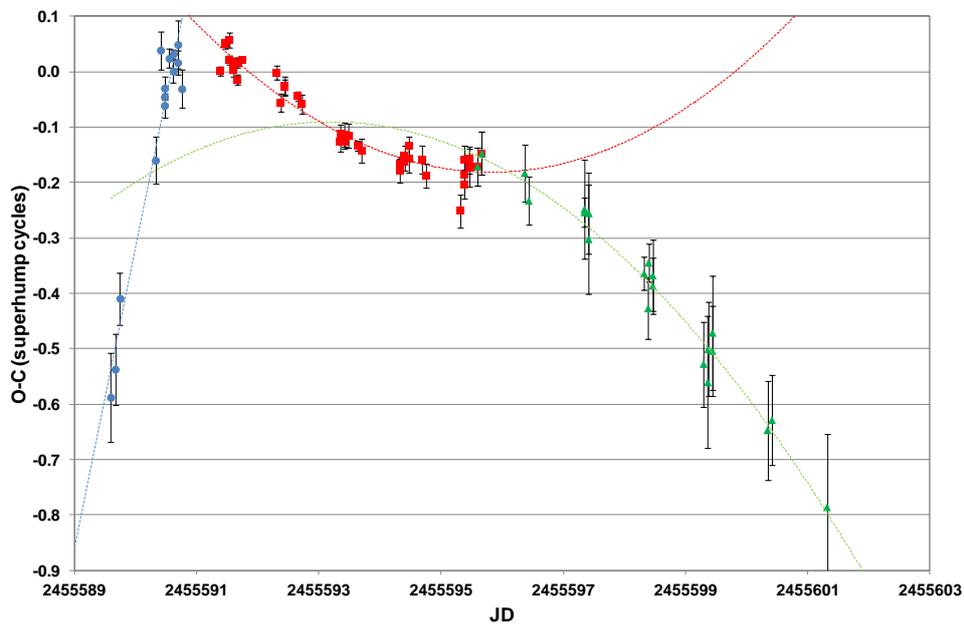
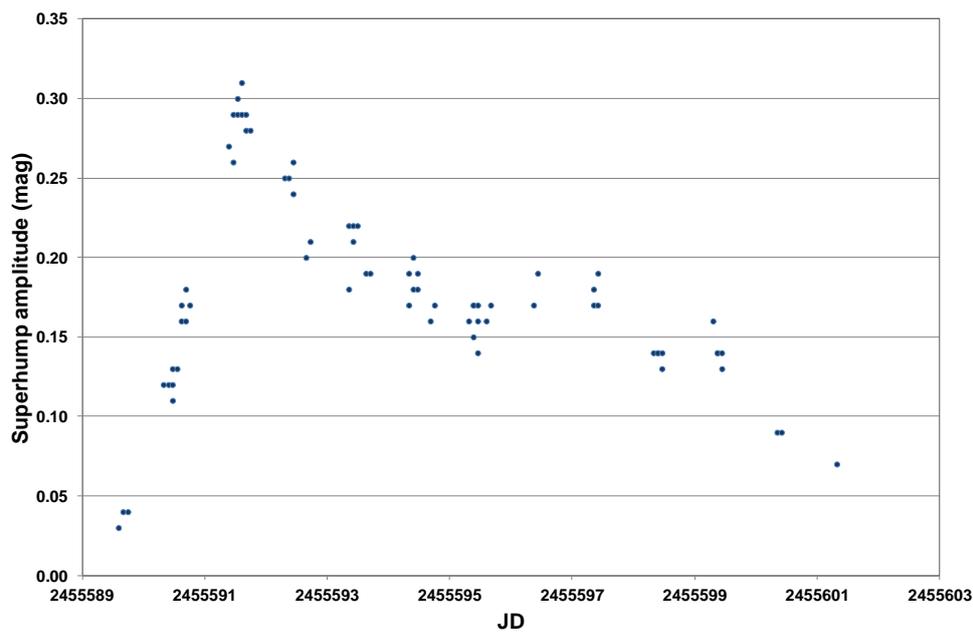

**Figure 1:** The 2011 outburst of V1212 Tau. Top: outburst light curve. Middle: O-C diagram. Bottom: superhump amplitude. See text for an explanation of the linear and quadratic fits on the O-C diagram

*Accepted for publication in the Journal of the British Astronomical Association*

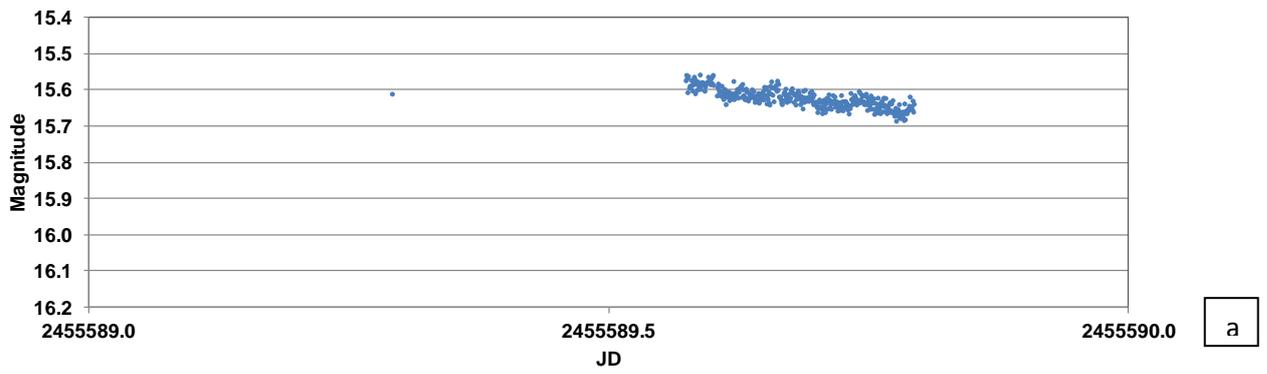

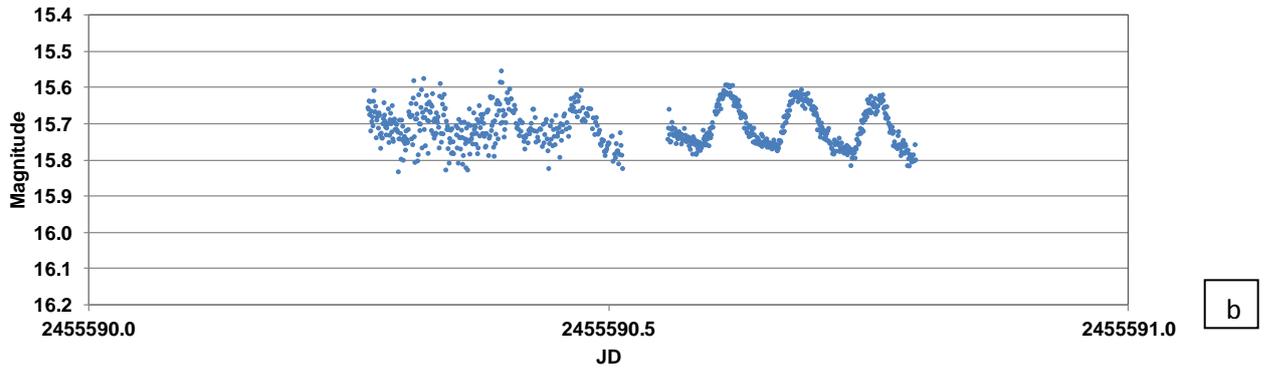

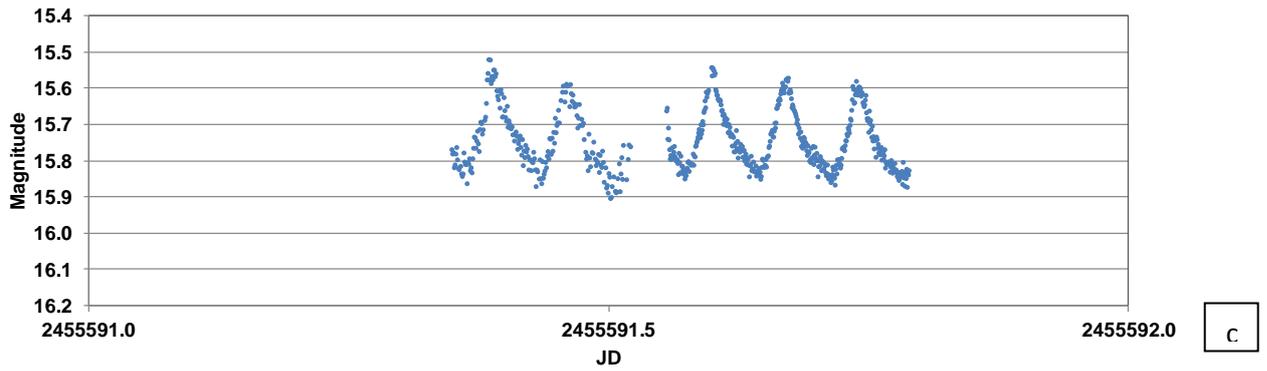

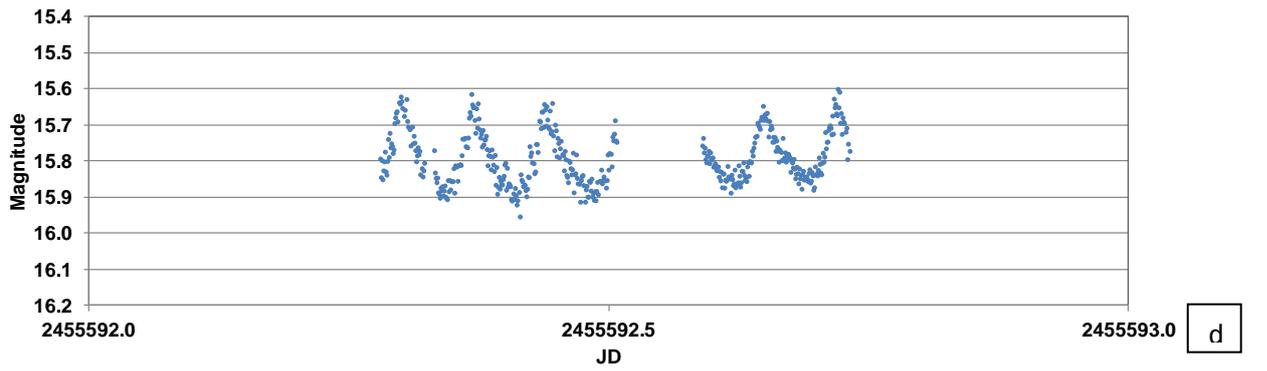

*Accepted for publication in the Journal of the British Astronomical Association*

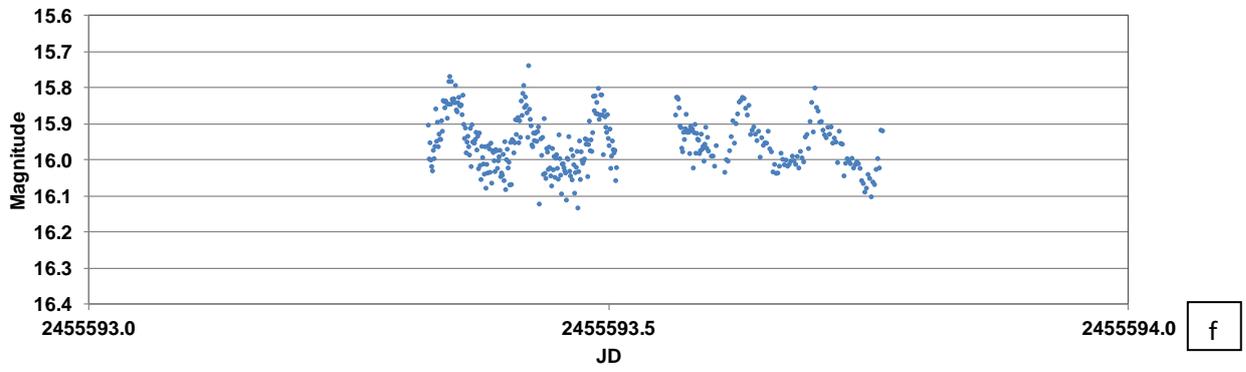

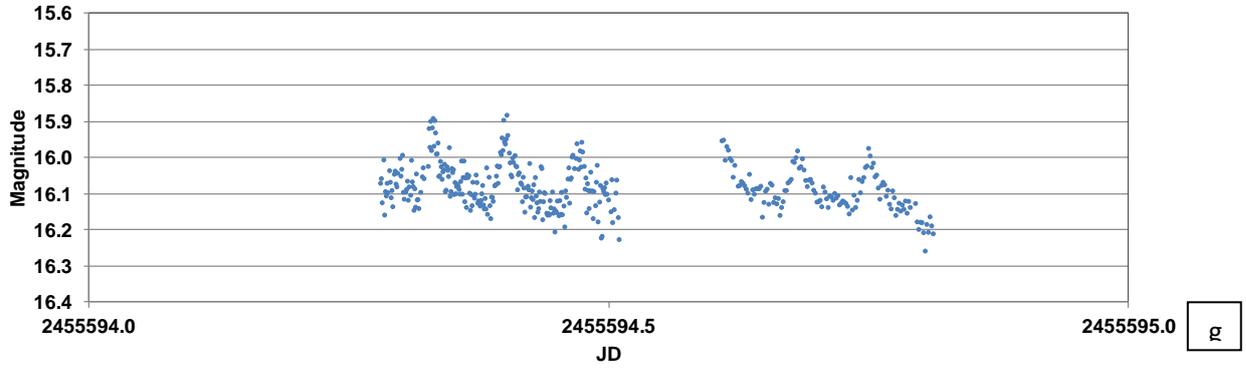

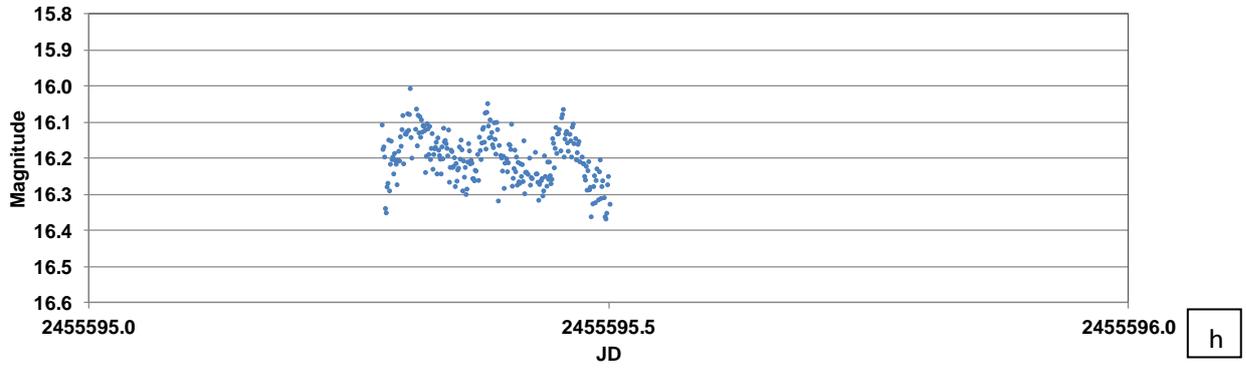

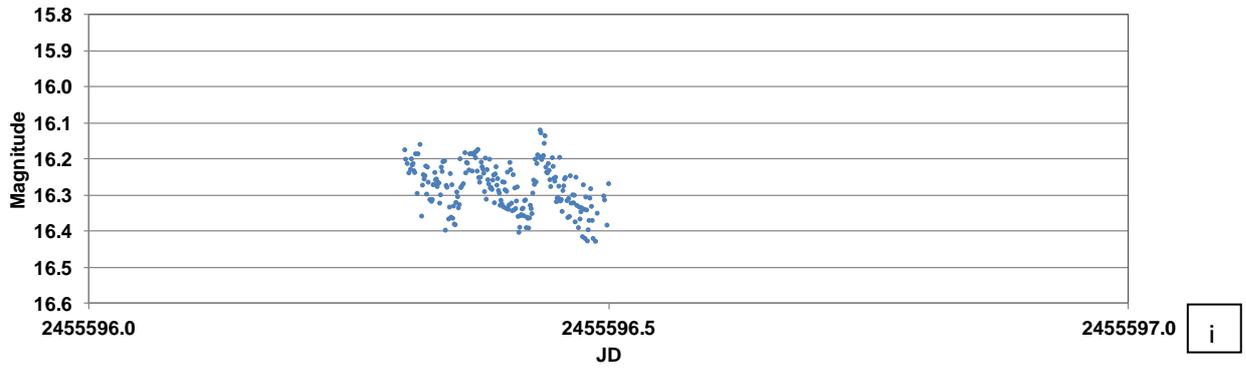

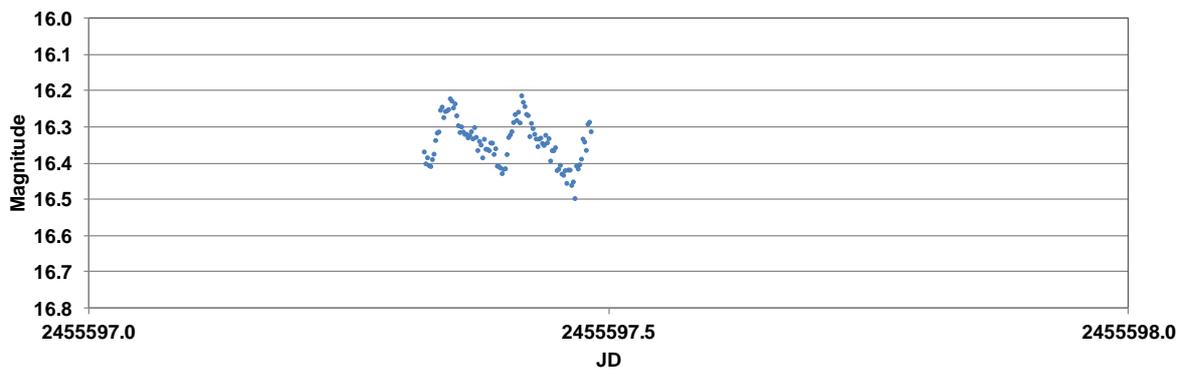

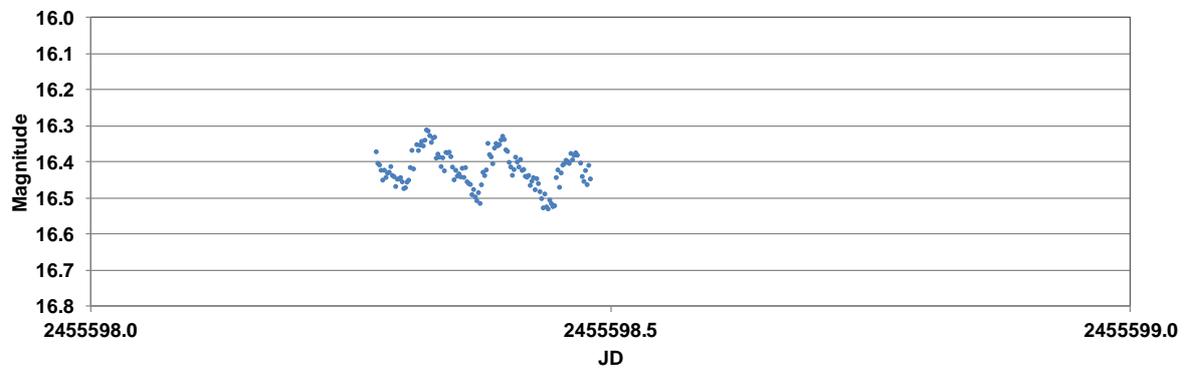
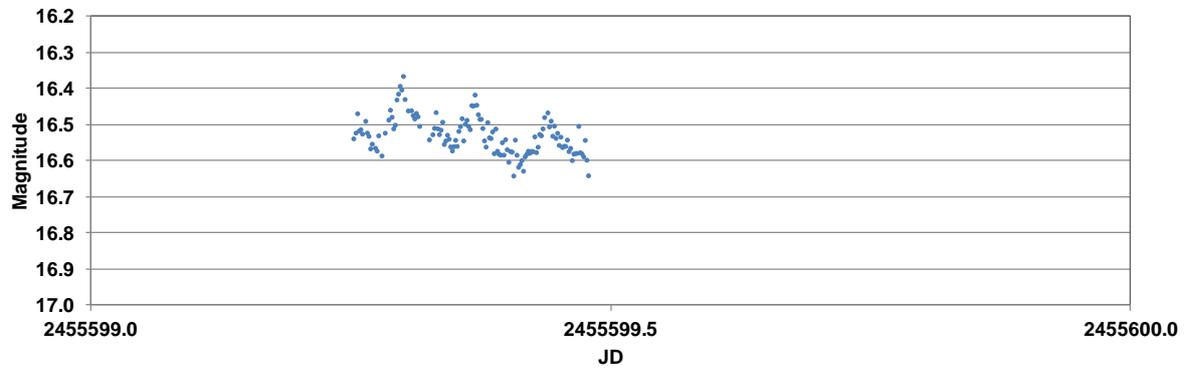
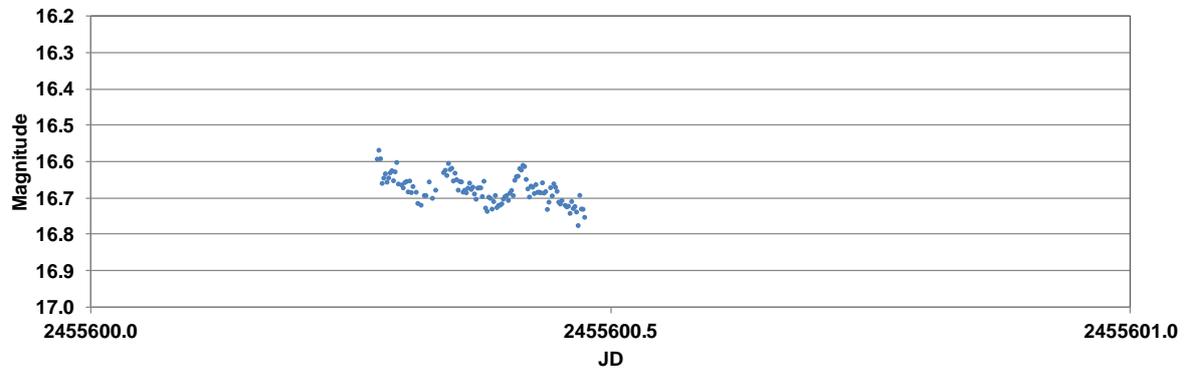

**Figure 2: Time resolved photometry during the outburst**

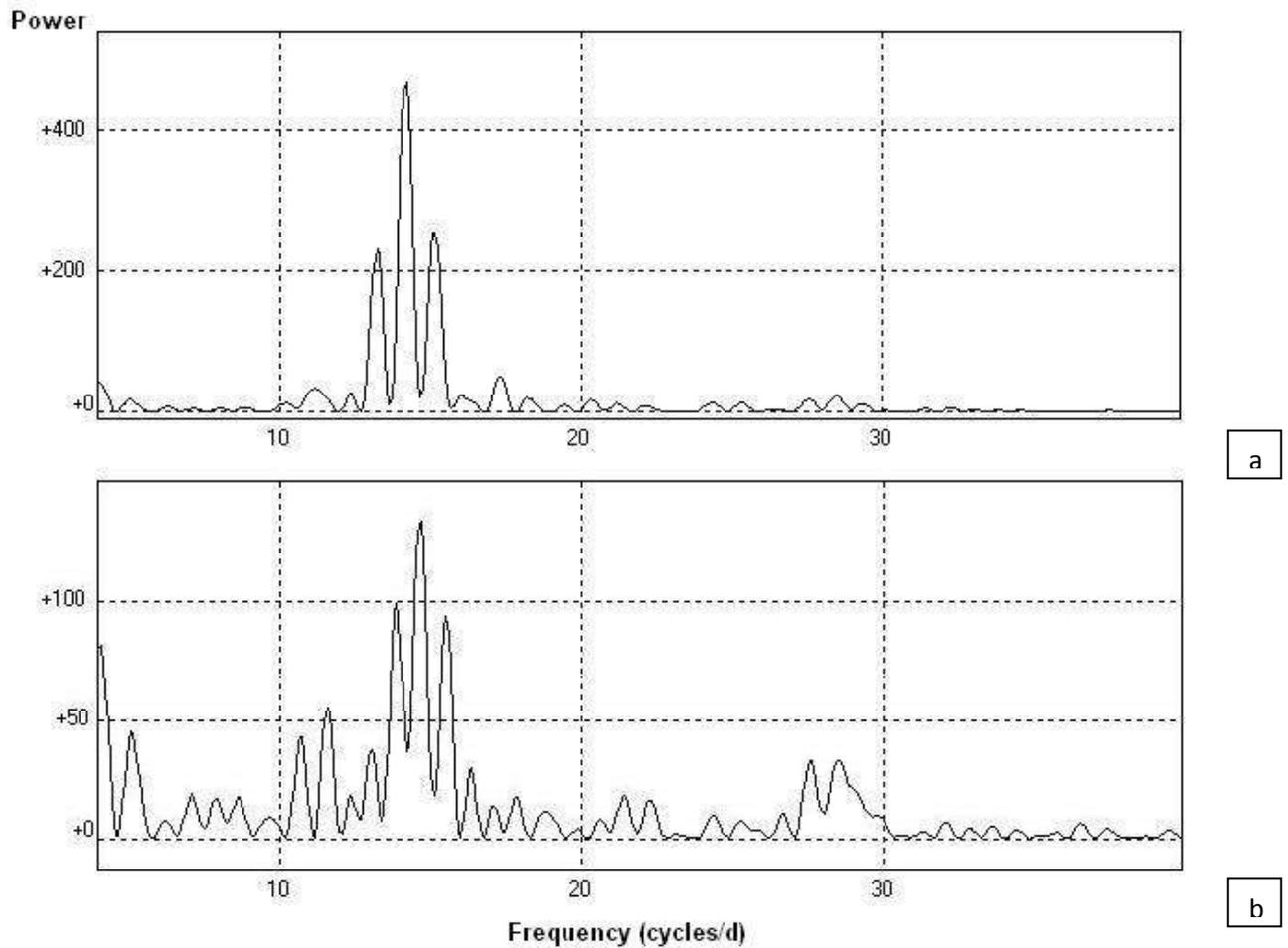

**Figure 3: Lomb-Scargle power spectra using data from JD 2455590 and 2455591**

(a) Power spectrum of combined data; (b) power spectrum after pre-whitening with the superhump signal

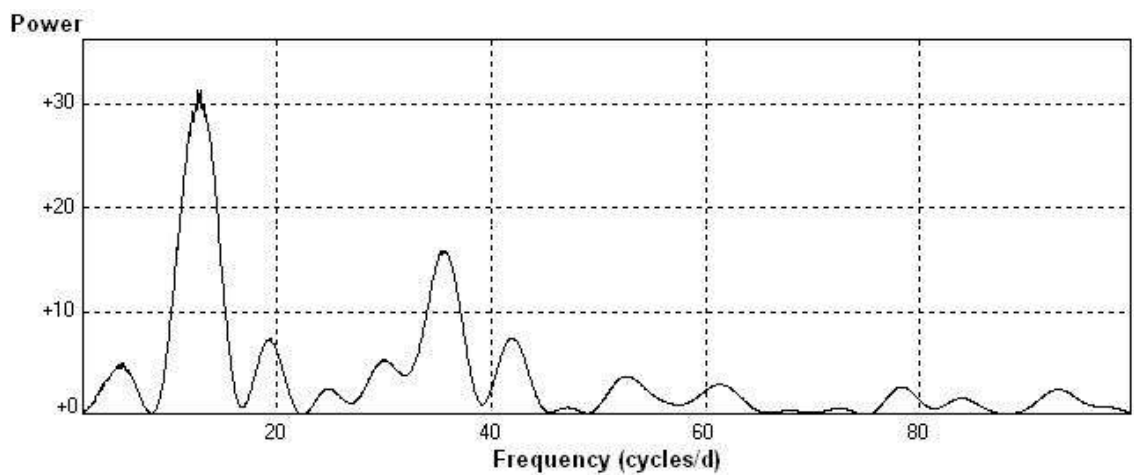

**Figure 4: Lomb-Scargle power spectrum using data from JD 2455589**